\documentclass[prl,twocolumn,showpacs,preprintnumbers,amsmath,amssymb,floatfix]{revtex4}
\usepackage{graphicx}% Include figure files
\usepackage{dcolumn}% Align table columns on decimal point
\usepackage{bm}% bold math
\usepackage{longtable}
\begin{document}

\title{ What drives the insulating state in ultrathin films of SrRuO$_3$?}
\author{Priya Mahadevan$^{1}$, F. Aryasetiawan$^{2}$, A. Janotti$^{3}$ and
T. Sasaki$^{4}$}
\affiliation{$^1$ S.N. Bose National Centre for Basic Sciences, JD-Block, Sector III, Salt Lake, Kolkata-700098, India. \\
$^2$ Graduate School of Advanced Integration Science, Chiba University, Chiba 263-8522, Japan \\
$^3$ Materials department, University of California, Santa Barbara, CA 93106-5050, USA \\
$^4$ National Institute of Materials Science, 1-2-1 Sengen, Tsukuba 305-0047, Japan
}
\date{\today}

\begin{abstract}
Using density functional calculations we have examined the evolution
of the electronic structure of SrRuO$_3$ films  grown on SrTiO$_3$ substrates
as a function of film thickness. At the ultrathin limit of two monolayers 
(RuO$_2$-terminated surface) the films are found to be at the brink of a spin-state transition which
drives the system to an antiferromagnetic and insulating state. Increasing the film
thickness to four monolayers, one finds the surprising result that two
entirely different solutions coexist. An antiferromagnetic insulating solution
coexists with a metallic solution corresponding to an antiferromagnetic 
surface and a ferromagnetic bulk. The electronic structure found at the
ultrathin limit persists for thicker films and an unusual result is predicted. 
Thicker films are found to be metallic as expected for the bulk, but
the magnetism does not directly evolve to the bulk ferromagnetic state.
The surface remains antiferromagnetic while the bulk exhibits ferromagnetic
ordering.
\end{abstract}

% Included pacs, please check 
\pacs{75.70.-i,71.15.-m,71.30.+h,75.50.Ee}
%75.70.-i 	         Magnetic properties of thin films, surfaces, and interfaces
%71.15.-m 	Methods of electronic structure calculations 
%71.30.+h 	Metal-insulator transitions and other electronic transitions 
% 75.50.Ee 	Antiferromagnetics 

\date{\today}
\maketitle

There has been considerable efforts in the search for alternate 
technologies to the current Si-based technology \cite{alt-tech}. 
An avenue of research that has been intensively explored is the use of transition-metal
perovskite oxides. The interest in these materials stems from their
multifunctional properties \cite{multifunctional}. Several of these materials 
exhibit different types of magnetic ordering, ferroelectricity, 
charge and orbital ordering, all resulting from a
strong interplay between spin, charge and orbital degrees of freedom. 
A set of interesting members of this class of compounds are those in 
which the magnetic ordering can be manipulated by an electric field, 
or the ferroelectric ordering can be manipulated by a magnetic field. 
Devices based on these materials have been envisioned
and this has fueled the study of oxide electronics \cite{oxide-electronics}. 
Metallic oxides form an integral part of the oxide-based device technologies, 
as they would form the connecting electrode material.

Among the perovskite oxides, SrRuO$_3$ is a promising material 
device applications, 
which apart from being a metal, although a poor one, also exhibits 
ferromagnetism up to 160 K \cite{srruo3-bulk}. An additional 
advantage of SrRuO$_3$ is that correlation effects should be
significantly weaker than in the 3$d$ transition-metal oxides. However,
when thin films of this material were epitaxially grown on SrTiO$_3$, 
films less than four monolayers thick were found to be insulating \cite{toyota,beasley}.
While the earlier studies reported the loss of 
ferromagnetic ordering in ultrathin films \cite{toyota}, recent studies have reported 
exchange bias effects which indicate the presence of antiferromagnetic regions \cite{beasley}.
All in all, the exact mechanism driving the metal-insulator transition 
in SrRuO$_3$ ultrathin films is still unknown. 
Could the metal-insulator transition be due to electron correlations, 
or could it be due to structural effects or extraneous experimental 
conditions as recently suggested \cite{spaldin}?
The question of whether electron correlations are important 
in these systems has been investigated experimentally, although there is
no consensus.  Some experimental studies suggest that electron-correlation effects 
are important even in the 4$d$ compounds and cannot be ignored \cite{kim,allen,geballe}. 
Other studies suggest that electron-correlation effects are much weaker and are not at all 
important in this class of materials \cite{maiti}. 
In order to address these issues, we performed first-principles density functional calculations 
of thin films of SrRuO$_3$ on a SrTiO$_3$ substrate, and examined the evolution of the 
electronic structure of the films with respect to the film thickness. 
We find that electron correlations at the level of GGA+$U$ must be 
included in order to explain the insulating state. 
The correlation effects serve the purpose of changing the subtle energy difference 
between the low-spin and the high-spin states. The transition into the high-spin 
state at the surface is accompanied by a strong Jahn-Teller-type distortion 
associated with a $d^4$ configuration with a partially filled $e_g$ states, which drives
the system to an insulating state. The choice of $U$ in GGA+$U$ 
calculations is usually arbitrary. In order to reduce this arbitrariness we
calculated $U$ from first-principles using a technique recently 
developed by one of the authors \cite{ferdi}.  
The magnitude of $U$ determined drives the thin films insulating.

SrRuO$_3$ is found to crystallize in an orthorhombic structure with 
four formula units per unit cell \cite{structure}. The Ru-O-Ru angles in the 
a-b plane are found to be around 168 degrees. SrTiO$_3$ is found to occur in a cubic structure.
The GGA optimized lattice constant of ferromagnetic cubic SrRuO$_3$ is 3.99 \AA,
while that of cubic SrTiO$_3$ is 3.94 \AA.  Therefore, there exists 
a small mismatch between cubic SrRuO$_3$ and SrTiO$_3$.
In order to simulate the ultrathin films of SrRuO$_3$ grown on SrTiO$_3$, 
we consider a 13 layer symmetric slab containing a central SrO layer, and alternating 
TiO$_2$/SrO layers along the (001) direction as shown in Fig. 1. 
The SrRuO$_3$/SrTiO$_3$ structures were constructed to represent SrRuO$_3$ films 
deposited on the SrTiO$_3$ substrate on a layer by layer mode.
In every case we have included 15 \AA~ of vacuum, and we have checked
that the results are converged with respect to the thickness of the vacuum layer.  
The in-plane lattice constant of the TiO$_2$ layers as well as the RuO$_2$ layers is 
set equal to the equilibrium lattice constant of SrTiO$_3$. 
As the Ru-O-Ru angles in bulk SrRuO$_3$ deviate from
180 degrees, we have allowed for a GdFeO$_3$ distortion of the RuO$_6$ octahedra
and optimised the structure to find the minimum energy solution. 
Full optimization of the atomic positions is performed. 
In addition, a ${\sqrt {2}}\times {\sqrt {2}}$ reconstruction of the (001) surface 
was considered, which allowed us to explore the possibility of antiferromagnetic ordering. 

The electronic structure of bulk as well as of thin films of SrRuO$_3$ 
were investigated using the plane-wave pseudopotential implementation 
of density functional theory and projector-augmented wave potentials in the 
VASP code \cite{vasp,paw}. 
The GGA approximation for the exchange-correlation functional was used.  
Electron-correlation effects at the Ru sites were included through 
the GGA+$U$ method within the Dudarev {\it et al.} formalism \cite{dudarev}, 
where $U$ represents the onsite Coulomb interaction strength and $J$ represents the 
intra-atomic exchange. 
We used the values of $U=2.5$ and $J$=0.4 eV for the Ru 4$d$ states,
which were determined for bulk  SrRuO$_3$ by using a 
recently proposed scheme based on the random-phase approximation \cite{ferdi}.
Although for the surface the $U$ may be larger due to the reduced
screening, we expect the qualitative aspects of the results to remain the same. 
A special k-point mesh of $6\times6\times6$ was used for integrations
over the the Brillouin zone of the bulk, while a mesh of $4\times4\times6$ 
was used for the slab calculations. A denser mesh of $8\times8\times4$ was
used in the density of states calculations for the slabs. 
A cutoff energy of  250 eV was used for the plane wave basis set.

{\it How does the electronic structure of bulk SrRuO$_3$ change in GGA+$U$?}.
We first examine the electronic structure of bulk SrRuO$_3$ as given by the GGA and GGA+$U$ functionals. 
The comparison between Ru 4$d$ partial density of states (PDOS) for $U$=0 (GGA) 
and $U=2.5$ eV, $J=0.4$ eV is shown  in Fig.~2.
The Fermi energy is used as reference by setting it to zero.
We note that there is an increase in the exchange splitting for the finite $U$ calculations. In order to 
examine the origin of the enhanced exchange splitting we plot the up- and
down-spin partial density of states in the insets of panels (a) and (b) in Fig.~2.
The up-spin density of states which were contributing at the 
Fermi level now move deeper into the valence band and the system becomes 
half-metallic as discussed earlier in the literature by Jeng et al. \cite{srruo3-oo}. In the down-spin channel we find that the effect of $U$
is minimal. Hence the increased exchange splitting comes from the
movement of up-spin states.
Gross features of the partial density of states are otherwise similar and have been 
previously discussed in the literature \cite{singh}.  For completeness, we  
repeat the discussion here in order to contrast with the surface which shows a very unusual electronic structure.

Ru in SrRuO$_3$ has a formal $d^4$ configuration. As the 
crystal-field splitting is quite large, the electrons occupy the $t_{2g}$ down-spin states after
occupying the $t_{2g}$ up-spin states. The system appears to be at the 
brink of a transition to a half-metallic state with a weak shoulder 
corresponding to the up-spin states at the Fermi level. It has been pointed
out earlier that a larger value of $U$ drives the system into a
half-metallic state \cite{srruo3-oo}. 
The moment on the Ru atom calculated by taking a 
sphere of radius 0.9 \AA~is 1 bohr magneton, significantly reduced from
the ionic value. 
This difference can be explained by the sizeable moments residing on the oxygen atoms.

Having understood the electronic structure of bulk SrRuO$_3$ we proceed by
examining the electronic structure of one atomic layer of RuO$_2$ grown
on the SrTiO$_3$ substrate. 
Before discussing the results from the calculations, we first speculate
on what to expect. Strong crystal-field anisotropies at the surface are
expected to result in a level ordering of doubly degenerate $d_{yz}$
and $d_{xz}$ levels followed by $d_{xy}$ levels. Thus if the system 
undergoes a transition to a nonmagnetic state the four electrons would 
go into the $d_{yz}$, $d_{xz}$ levels. This would explain the origin of the
insulating ground state. With this 
simple model in mind, we examined various possible solutions. 
Along with the ferromagnetic configuration, we considered nonmagnetic as well as antiferromagnetic configurations.
For $U$=0, the ferromagnetic configuration is the ground state, being more 
stable than the nonmagnetic configuration by  24 meV per RuO$_2$ unit, whereas 
the antiferromagnetic solution is difficult to converge to.
However, for $U$=2.5 and a $J$=0.4 eV on Ru, we find that the antiferromagnetic state is 
more stable by 49 meV per RuO$_2$ unit. Therefore, in every case, {\it
magnetism survives at this ultrathin film limit.}  Most importantly,
the transition to the antiferromagnetic state is accompanied by a change in the electronic
state from metal to insulator.

{\it What is the origin of the insulating state?} 
While it is clear from our results that the insulating state is linked to an antiferromagnetic order, 
the origin of the insulating state is puzzling, and we rely on 
an analysis of the projected density of states to explain it.
The orbital- and spin-projected density of states of Ru shown in Fig.~3,
and the corresponding schematic energy-level diagram shown in Fig.~4.
In the up-spin channel we find that the $d_{xy}$, $d_{yz}$ and $d_{xz}$ states
have a significant weight in an energy window near $-5$ eV as well as near 0 eV,
and the exact ordering of these $t_{2g}$-derived levels is not very clear.
However, when we examine the contributions of the down-spin counterparts,
the ordering becomes clear. The states with $d_{xz}$ and $d_{yz}$ character
are at lower energies compared to the state with $d_{xy}$ character.

The origin of this splitting of the $t_{2g}$ levels is clear when one examines
the bond lengths in the RuO$_5$ unit at the surface. The in-plane 
bond lengths are equal to $\sim$1.95-1.96 \AA, while the apical oxygen
bond length is $\sim$2.13 \AA. Hence crystal field effects determine
the ordering of the t$_{2g}$ states, with the {$d_{yz}$ and $d_{xz}$} levels followed by $d_{xy}$
in each spin channel. This ordering is depicted schematically in the
left panel of Fig.~4. These levels interact with the corresponding symmetry
levels on the oxygen atoms forming bonding and antibonding states. We
illustrate this interaction explicitly for the up-spin $t_{2g}$-derived
levels in Fig.~3.  As the charge transfer energy between the up-spin $t_{2g}$
and the oxygen atoms is small, the bonding states arising from this interaction 
have significant Ru $d$ character. The hopping interactions however should
be strongest for the in-plane $d_{xy}$ orbital. This results in
an inversion of the ordering of the bonding levels generated as a result of
the interaction.

The strong crystal field that is present at the surface also induces a huge
splitting of the $e_g$-derived levels. The $d_{3z^2-r^2}$ states,
which were unoccupied in the bulk, are now occupied, while the up-spin
$d_{x^2-y^2}$ state lies above the down-spin $d_{xy}$, $d_{yz}$ and $d_{xz}$
states. Hence an unusual ordering of levels emerges from the broken 
symmetry at the surface.

The next question we asked was what is the energetics  
involved in driving the unusual level ordering that one finds here.
The generation of the large crystal distortion would cost enormous
strain energy which we speculate must be compensated by the additional
Hund's intra-atomic exchange present as a result of the high-spin state on Ru
being favored. It should be noted that this crystal-field
distorted structure exists above a particular value of $U$ for the
antiferromagnetic state alone. The nonmagnetic and ferromagnetic 
solutions do not show such distortion at the two monolayers limit.

Experimentally it is found that the insulating state exists for four
atomic layers, so it is important to ask whether the insulating
state survives when we have an additional RuO$_2$ layer. 
This has direct correspondence with experiment as it corresponds
to four atomic layers on the exposed TiO$_2$ surface. 
The TiO$_2$ surface is usually the one exposed in the experiments 
on which the SrRuO$_3$ films are grown \cite{pvt-comm}. 
The Ru $d$-projected PDOS for the atoms at the surface layer 
as well as at the sub-surface layer are shown in Fig.~5. 
The system is still found to be insulating. Interestingly the surface Ru atoms
show a similar structural distortion as found in the single RuO$_2$ layer case.
However, the inner RuO$_2$ layer does not show the same distortion
but still is insulating. Thus it is the antiferromagnetic correlations
which set in as a result of the crystal distortion of the surface layer
which drives the sub-surface RuO$_2$ layer insulating.  

In order to probe the thickness at which we have an insulator-metal
transition, we investigated the effects of additional RuO$_2$ layers on the
electronic structure of the films.
Surprisingly, we find that the antiferromagnetic insulating solution is always lower than the
ferromagnetic metallic solution for 
even 8 monolayers.  This is far beyond the thickness for which a
metal-insulator transition has been observed in experiments. 
One might suggest that $U$=2.5 eV used in our calculations is too large for the ruthenates.
However, we note that the exchange bias effects that has been recently observed in
these systems \cite{beasley} offer us some hints to the solution of this puzzle.
{\it Could the experimental results be explained by an antiferromagnetic surface 
and a ferromagnetic bulk (AFMS-FMB) ?}
By examining this further we found that
the antiferromagnetic solution is actually degenerate with the AFMS-FMB state at the
four monolayer limit. The antiferromagnetic solution as we pointed 
out earlier is insulating, while the AFMS-FMB solution is metallic.
Thus, our results indicate that two electronically different states coexist, and that 
disorder would possibly pin one solution in one region. 
For thicker films of six or more monolayers we find the AFMS solution to be the most stable. 
The AFMS-FMB solution we find is the favored solution far away from the ultrathin limit, even for SrRuO$_3$ 
surfaces.

Overall, thin films of SrRuO$_3$ of 4 monolayer thickness or less are found to be 
insulating. Examining the electronic structure of ultrathin films with 
just one RuO$_2$ layer, we find that the Ru atom at the surface undergoes
a low-spin to high-spin transition. This rare occurrence of a high-spin
state in a 4$d$ oxide results in an antiferromagnetic state being stabilized
as the ground state. The antiferromagnetic ordering drives the system 
insulating. In the limit of 4 monolayers, the antiferromagnetic insulating
solution coexists with the solution corresponding to an antiferromagnetic
surface and a ferromagnetic bulk (AFMS-FMB).  For six-monolayers thick films, 
we find antiferromagnetic surface and the bulk ferromagnetic solution, 
which is metallic, to have lower energy. In other words, we find that
SrRuO$_3$ surfaces are antiferromagnetic, while the bulk is ferromagnetic.

Hence the electronic structure of systems in the ultrathin limit are very
different from the bulk. Crystal-field anisotropies largely determine
the properties of ultrathin films. In the case of SrRuO$_3$ we show that a metal to
insulator transition takes place as a function of thickness in addition
to a ferromagnet to an antiferromagnet transition.

PM thanks the Department of Science and Technology for financial support.

\renewcommand
\newpage

\begin{figure}
\includegraphics[width=5.5in,angle=270]{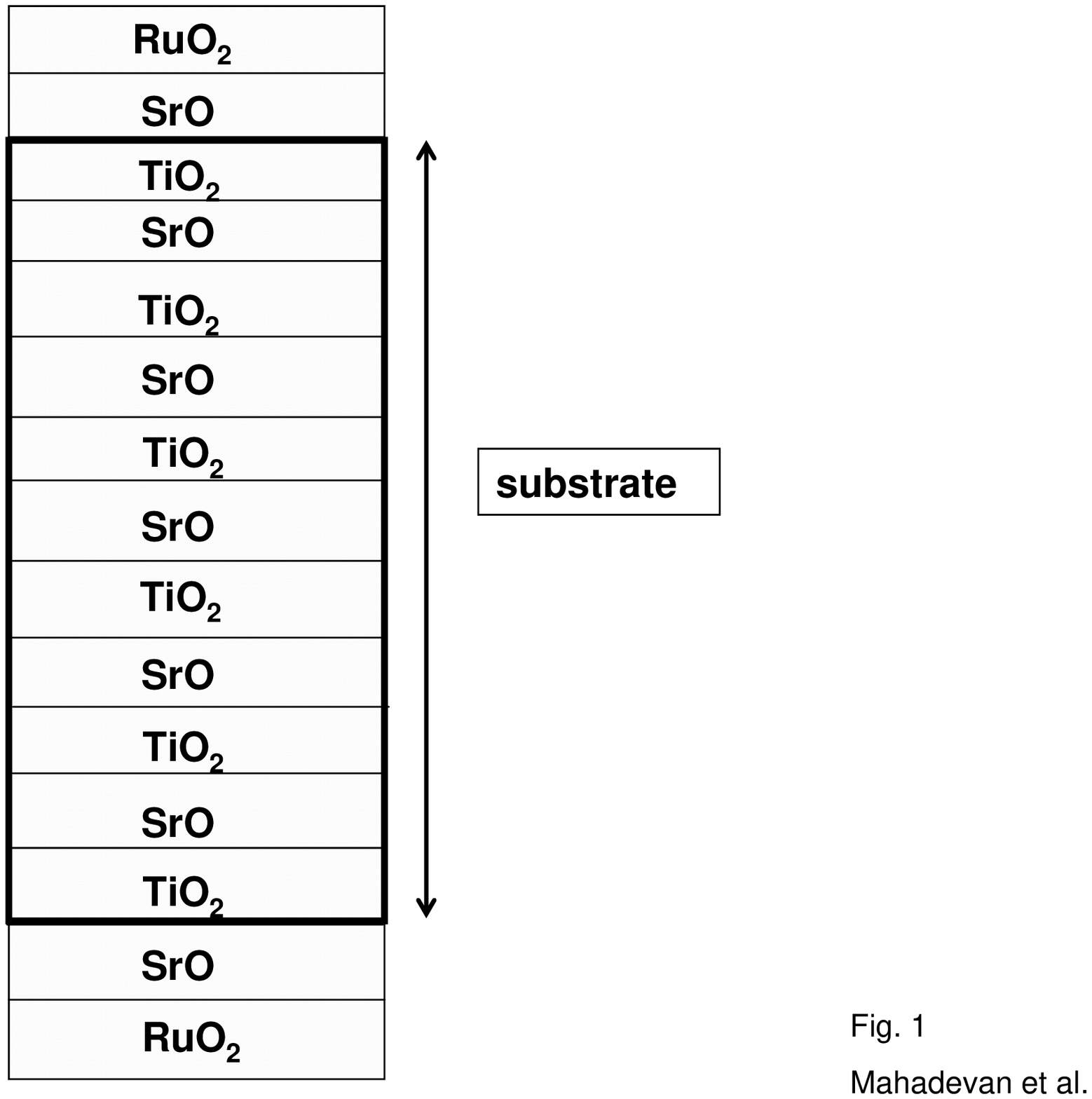}
\caption{ A schematic diagram of the slab geometry that was considered.
The substrate was fixed at 11 monolayers on which the SrRuO$_3$ films
were grown. The present figure corresponds to 2 monolayers of SrRuO$_3$.
}
\end{figure}
\begin{figure}
\includegraphics[width=5.5in]{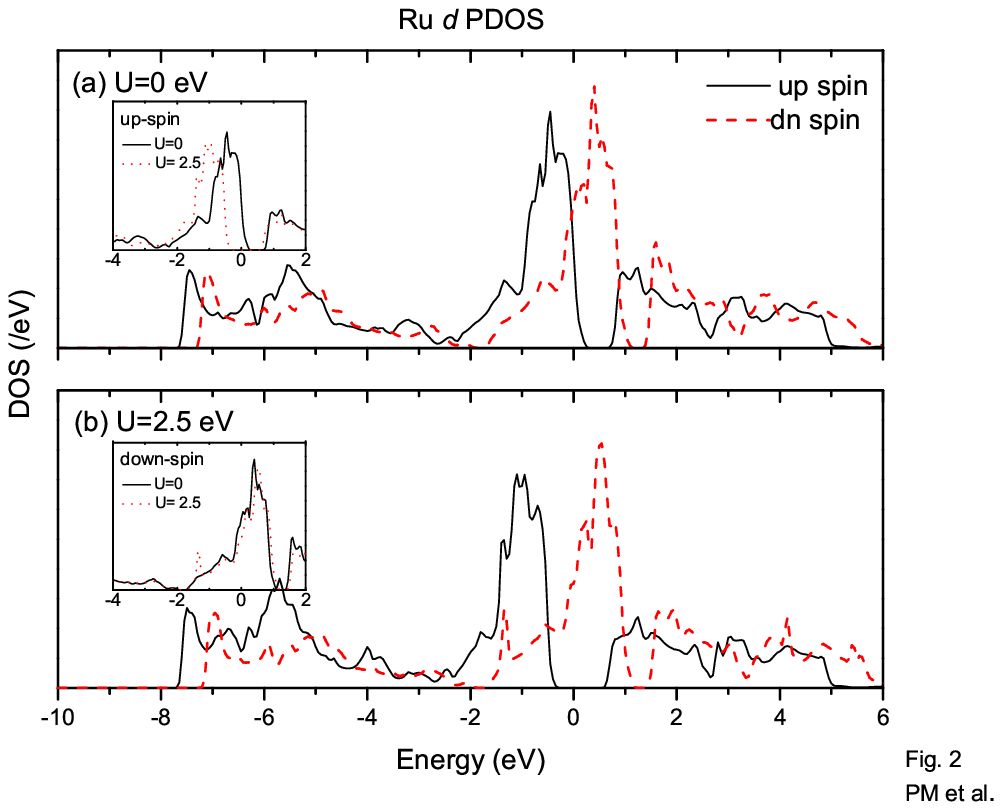}
\caption{ The up (solid line) and down (dashed line) spin
Ru $d$ partial density of states for bulk SrRuO$_3$  in the ferromagnetic
state.The zero of energy corresponds to the Fermi energy.
}
\end{figure}
\begin{figure}
\includegraphics[width=5.5in]{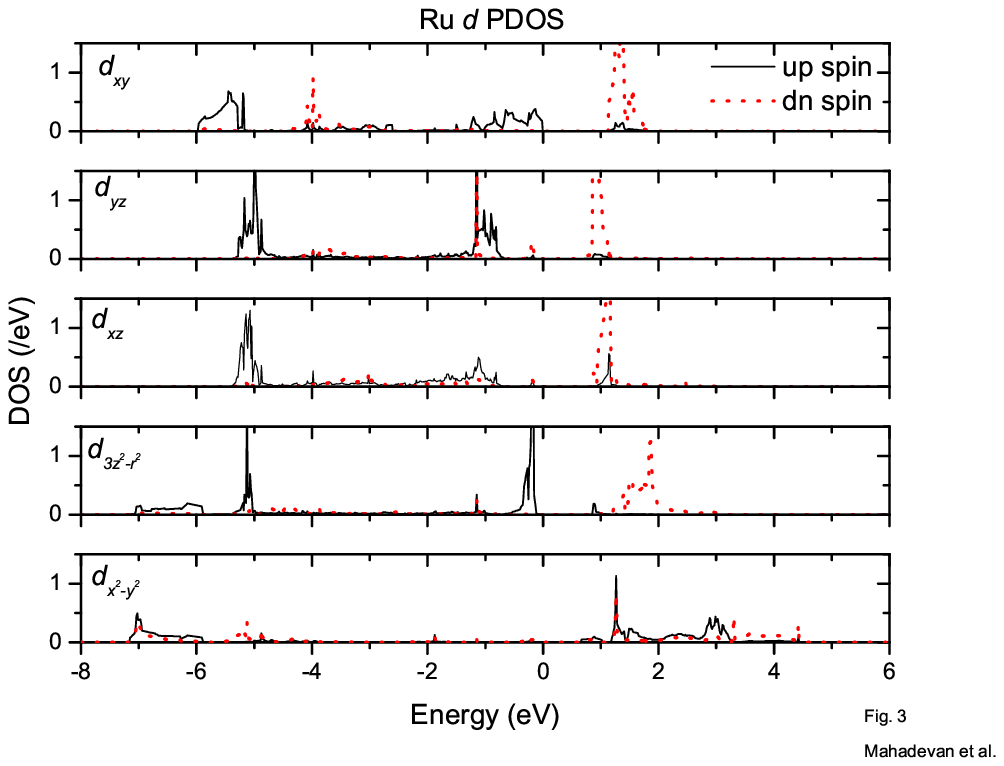}
\caption{ The up (solid line) and down (dashed line) spin
orbital projected 
Ru $d$ partial density of states for the surface Ru atom for 2 monolayers
of SrRuO$_3$ grown on SrTiO$_3$
calculated for the antiferromagnetic state. The zero of energy corresponds to the top of the valence band.
}
\end{figure}
\begin{figure}
\includegraphics[width=5.5in,angle=270]{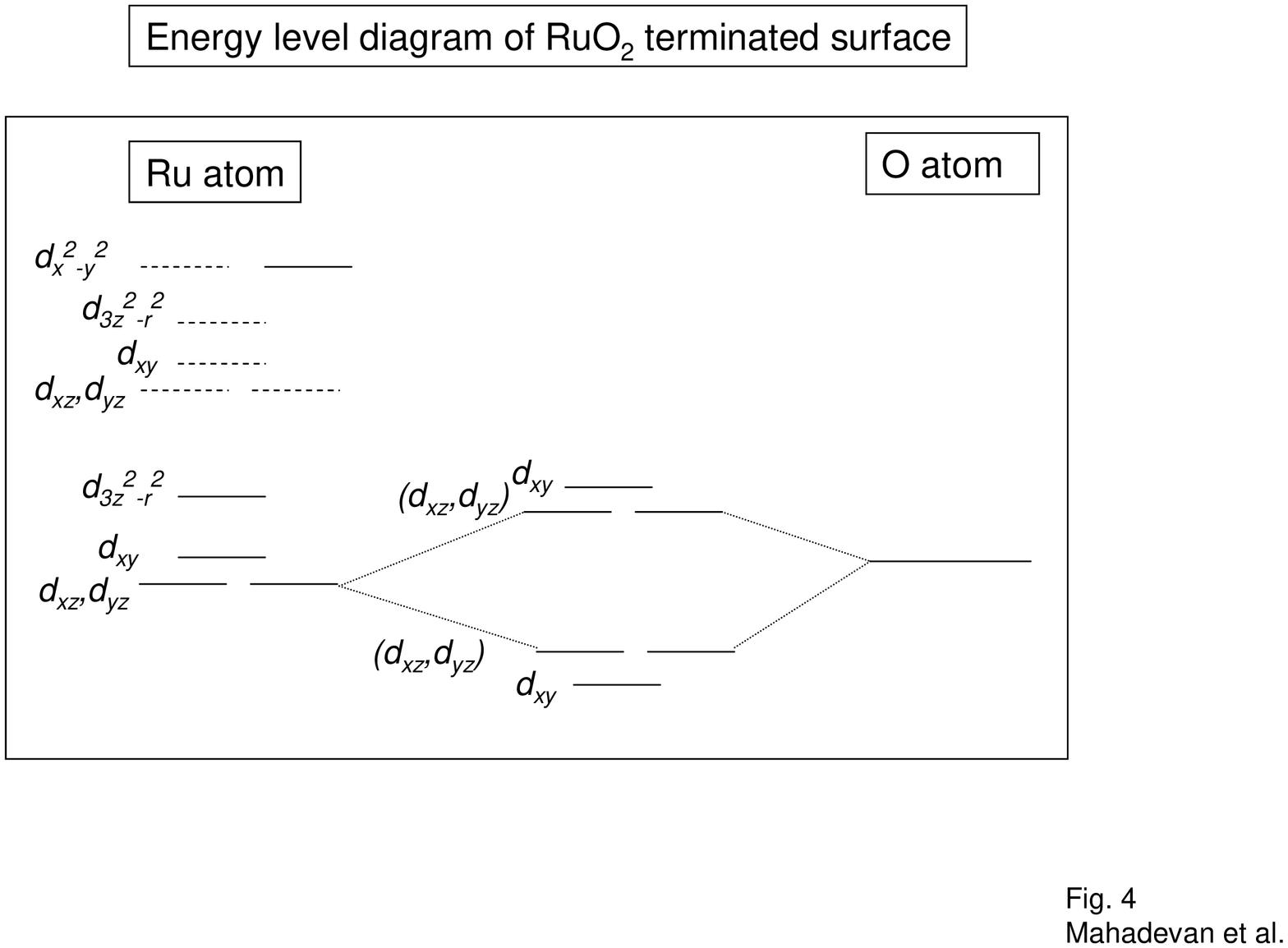}
\caption{ The up (solid line) and down-spin (dashed line) crystal field
split Ru $d$ levels for the surface atom are shown schematically in the 
left panel. The interaction of these levels with the O $p$ orbitals (in the 
right panel) and the ensuing ordering of the up-spin $t_{2g}$-derived
levels are shown in the central panel.}
\end{figure}

\begin{figure}
\includegraphics[width=5.5in]{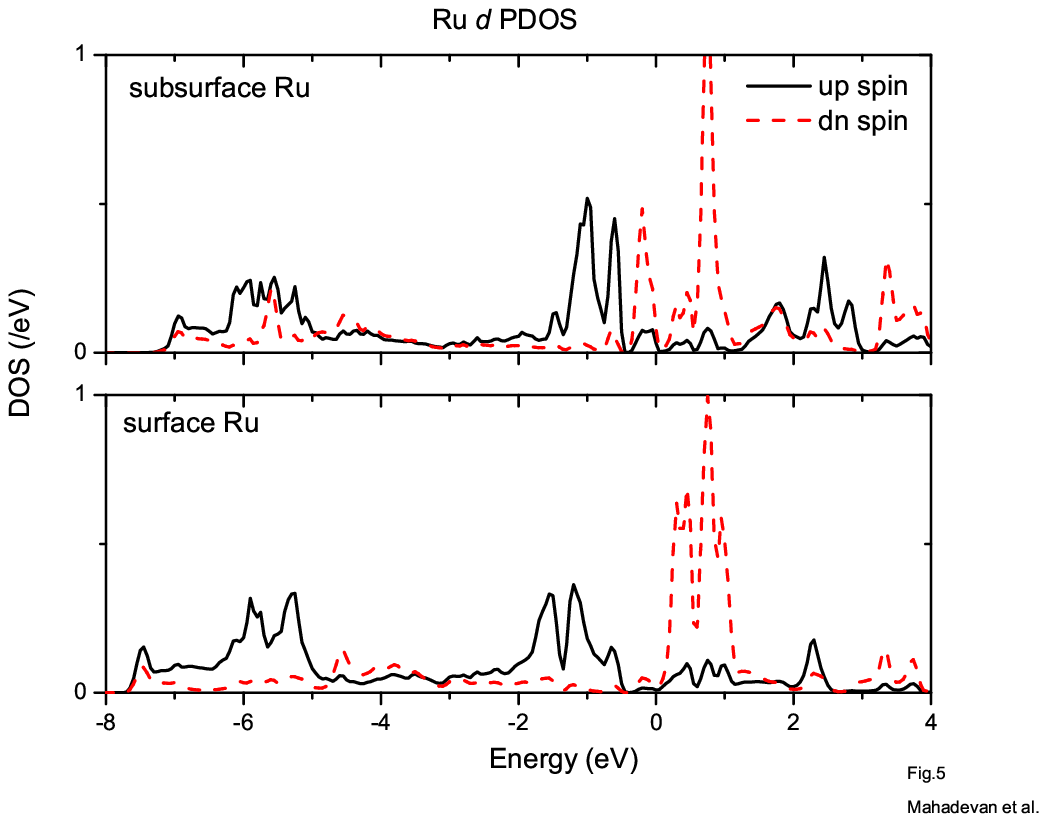}
\caption{ The up (solid line) and down (dashed line) 
spin Ru $d$ partial density of states for 4 monolayers 
of SrRuO$_3$ grown on SrTiO$_3$ for (a) sub-surface Ru atom as well
as for (b) the surface Ru atom are shown for the
antiferromagnetic state. The zero of energy corresponds to the 
top of the valence band.}
\end{figure}

\end{document}